\documentclass[reprint,superscriptaddress,showkeys,showpacs,amsmath,amssymb,aps,prl]{revtex4-2}

\usepackage{graphicx}
\usepackage{dcolumn}
\usepackage{bm}
\usepackage{color}
\definecolor{darkblue}{rgb}{0,0,.6}
\usepackage[colorlinks=true, citecolor=darkblue, linkcolor=darkblue, menucolor=darkblue, urlcolor=darkblue]{hyperref}
\usepackage{upgreek}

\newcommand{\RuCl}{$\alpha$-RuCl$_3$}
\newcommand{\GK}{$(\xi\xi0)$}
\newcommand{\GM}{$(\xi00)$}
\newcommand{\GA}{$(00\xi)$}
\begin{document}

\preprint{arxiv:blah.blah}

\title{Acoustic phonon dispersion of \texorpdfstring{\RuCl{}}{alpha-RuCl3}}

\author{Blair~W.~Lebert}
\author{Subin~Kim}
\affiliation{Department of Physics, University of Toronto, Toronto, Ontario, M5S 1A7, Canada}
\author{Danil~A.~Prishchenko}
\affiliation{Department of Theoretical Physics and Applied Mathematics, Institute of Physics and Technology, Ural Federal University, Mira Str. 19, 620002 Ekaterinburg, Russia}

\author{Alexander~A.~Tsirlin}
\affiliation{Department of Theoretical Physics and Applied Mathematics, Institute of Physics and Technology, Ural Federal University, Mira Str. 19, 620002 Ekaterinburg, Russia}
\affiliation{Experimental Physics VI, Center for Electronic Correlations and Magnetism, Institute of Physics, University of Augsburg, 86135 Augsburg, Germany}

\author{Ayman~H.~Said}
\author{Ahmet~Alatas}
\affiliation{Advanced Photon Source, Argonne National Laboratory, Lemont, IL 60439, USA}
\author{Young-June~Kim}
\affiliation{Department of Physics, University of Toronto, Toronto, Ontario, M5S 1A7, Canada}
\email{youngjune.kim@utoronto.ca}

\date{\today}

\begin{abstract}
Acoustic phonons have recently been posited as playing an integral role in explaining the half-quantized thermal Hall effect in \RuCl{}. Therefore, we present much needed inelastic x-ray scattering measurements of its acoustic phonon dispersion, along with calculations using the frozen-phonon method. We also discuss a temperature study which conclusively shows a first-order structural transition to a non-$C2/m$ space group at low temperature. Together these results are an important backbone for future theoretical and experimental studies of \RuCl{}.
\end{abstract}

%\keywords{}
%\pacs{}
\maketitle

Spin-orbit assisted Mott insulators \cite{witczak-krempaCorrelatedQuantumPhenomena2014} are a subject of great interest in the search for topologically ordered quantum spin liquids (QSLs), which offer a route towards fault-tolerant quantum computation \cite{kitaevFaulttolerantQuantumComputation2003,nayakNonAbelianAnyonsTopological2008}. In particular, Kitaev's honeycomb model \cite{kitaevAnyonsExactlySolved2006} has been an incredibly fruitful domain, with many Kitaev QSLs candidates being synthesized \cite{rauSpinOrbitPhysicsGiving2016,trebstKitaevMaterials2017,hermannsPhysicsKitaevModel2018,takagiConceptRealizationKitaev2019} following the pioneering proposals of Jackeli and Khaliullin \cite{jackeliMottInsulatorsStrong2009,chaloupkaKitaevHeisenbergModelHoneycomb2010,chaloupkaZigzagMagneticOrder2013}. \RuCl{} is one of the more promising candidates \cite{plumbRuClSpinorbitAssisted2014,majumderAnisotropicRuMagnetism2015,sandilandsScatteringContinuumPossible2015,searsMagneticOrderRuCl2015,banerjeeProximateKitaevQuantum2016,johnsonMonoclinicCrystalStructure2015,caoLowtemperatureCrystalMagnetic2016,parkEmergenceIsotropicKitaev2016,sandilandsSpinorbitExcitationsElectronic2016,banerjeeNeutronScatteringProximate2017,doMajoranaFermionsKitaev2017,lebertResonantInelasticXray2020,suzukiProximateFerromagneticState2021,liGiantPhononAnomalies2021}, especially since an in-plane magnetic field of ${\approx}$8~T suppresses its low-temperature zigzag magnetic order \cite{kubotaSuccessiveMagneticPhase2015,baekEvidenceFieldInducedQuantum2017,leahyAnomalousThermalConductivity2017,searsPhaseDiagramRuCl2017,wolterFieldinducedQuantumCriticality2017,banerjeeExcitationsFieldinducedQuantum2018,hentrichUnusualPhononHeat2018a,jansaObservationTwoTypes2018,balzFiniteFieldRegime2019,widmannThermodynamicEvidenceFractionalized2019,schonemannThermalMagnetoelasticProperties2020a,bachusThermodynamicPerspectiveFieldInduced2020,bachusAngledependentThermodynamicsRu2021a}. Recently in this intermediate field regime, half-integer quantized thermal Hall conductivity \cite{kasaharaMajoranaQuantizationHalfinteger2018,yamashitaSampleDependenceHalfinteger2020,yokoiHalfintegerQuantizedAnomalous2021,bruinRobustnessThermalHall2021} and quantum oscillations of the thermal conductivity \cite{czajkaOscillationsThermalConductivity2021} were discovered. These results provide strong evidence of chiral Majorana edge currents emerging from a gapped Kitaev-like QSL.

Thermal Hall conductivity is the tool of choice for probing these charge-neutral excitations since they cannot be probed by electrical Hall conductivity. However, unlike electrical transport, thermal transport is dominated by bulk acoustic phonons. In \RuCl{} the thermal Hall angle is small, $\kappa_{xy}/\kappa_{xx} \approx 10^{-3}$ \cite{kasaharaMajoranaQuantizationHalfinteger2018,yamashitaSampleDependenceHalfinteger2020,yokoiHalfintegerQuantizedAnomalous2021,bruinRobustnessThermalHall2021}. Such a small Hall angle can disrupt quantization in the electrical Hall effect. A similar situation could be found in \RuCl{} due to bulk phonons coupling with chiral edge modes. However, recent papers from \citet{yeQuantizationThermalHall2018} \& \citet{vinkler-avivApproximatelyQuantizedThermal2018} propose that this edge-bulk coupling not only allows approximate quantization for sufficiently large rectangular samples but is actually necessary since thermal leads couple primarily through bulk phonons rather than edge modes. These studies highlight the importance of the lattice and its dynamics in \RuCl{}. The acoustic phonons of \RuCl{} have been measured with inelastic x-ray scattering (IXS) \cite{liGiantPhononAnomalies2021, liDivergenceMajoranaPhononScattering2021}, however these studies did not systematically study all acoustic phonons along all high-symmetry directions. As well, inelastic neutron scattering (INS) has measured phonons up to 50~meV, but the relatively coarse resolution of this study meant that the acoustic phonon velocities were determined using first-principles calculations fit to dataset rather than directly fitting acoustic phonon branches \cite{muRoleThirdDimension2021}.

%%%%%%%%%%%%%%%%%%%%%%%%%%%%%%%%%%%%%%%%%%%%%%%%%%%%%%%%%%

\begin{figure}[t]
\includegraphics{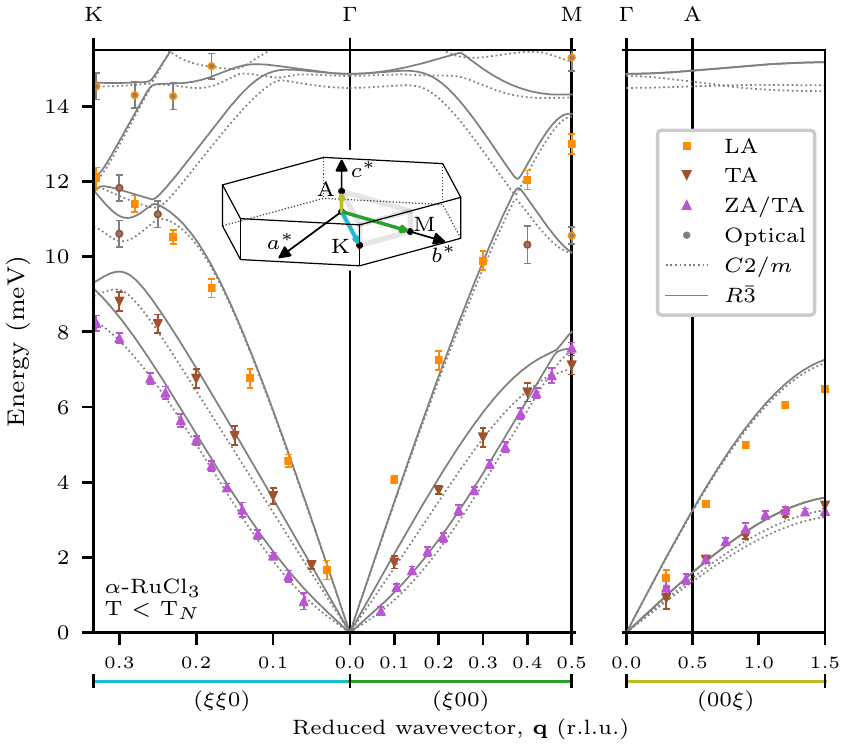}
\caption{\label{fig:dispersion} 
Inelastic x-ray scattering measurement of phonons in \RuCl{}. All data was collected at 5.7~K, except for the ZA modes which were measured at 4.1~K. The optical phonons are outlined based on measurement geometry. Gray lines indicate calculations where force constants are calculated with $R\bar3$ (solid) and $C2/m$ (dotted) space groups. Inset: Brillouin zone showing measurement paths.}
\end{figure}

%%%%%%%%%%%%%%%%%%%%%%%%%%%%%%%%%%%%%%%%%%%%%%%%%%%%%%%%%%

Regarding the lattice structure, recent papers agree that the ambient temperature space group is $C2/m$ \cite{johnsonMonoclinicCrystalStructure2015, caoLowtemperatureCrystalMagnetic2016,parkEmergenceIsotropicKitaev2016,lebertResonantInelasticXray2020,muRoleThirdDimension2021} and not $P3_112$ as initially reported \cite{stroganov1957}, however the low temperature structure is still debated. Many probes find a hysteretic transition at ${\approx}$145~K \cite{kubotaSuccessiveMagneticPhase2015,searsNeutronXrayDiffraction2017,parkEmergenceIsotropicKitaev2016, widmannThermodynamicEvidenceFractionalized2019, gassFieldinducedTransitionsKitaev2020, reschkeElectronicPhononExcitations2017, reschkeSubgapOpticalResponse2018, glamazdaRelationKitaevMagnetism2017, wangMagneticExcitationsContinuum2017, heUniaxialHydrostaticPressure2018} which \citet{parkEmergenceIsotropicKitaev2016} associates with a structural transition to an $R\bar3$ space group. On the other hand, \citet{caoLowtemperatureCrystalMagnetic2016} and \citet{johnsonMonoclinicCrystalStructure2015} both do not observe a symmetry change. Recently, neutron diffraction measurements by \citet{muRoleThirdDimension2021} and a reanalysis \cite{muRoleThirdDimension2021} by \citet{caoLowtemperatureCrystalMagnetic2016} both support a $R\bar3$ space group at low temperature. Therefore, a consistent picture of the low-temperature space group is emerging, however peak splitting in neutron diffraction still suggests a space group with lower symmetry \cite{muRoleThirdDimension2021}. The main difference between the $R\bar3$ and $C2/m$ structures is the stacking order of the RuCl$_3$ layers. Stacking order is a recurrent theme in \RuCl{} and poor-quality samples have been shown to have a large number of stacking faults, leading to partial magnetic order at 14~K \cite{kimSampleQualityRuCl32021}. With regards to the thermal Hall effect in \RuCl{}, sample quality is paramount since quantization only occurs in high-quality samples with a single magnetic transition, low stacking faults, and high $\kappa_{xx}$ \cite{yamashitaSampleDependenceHalfinteger2020,yokoiHalfintegerQuantizedAnomalous2021}.

In this Letter, we systematically measure the acoustic phonon dispersion of high-quality single crystals of \RuCl{} with high-resolution IXS. Our experimental results are modelled extremely well using the frozen-phonon method with forces calculated from DFT+$U$+SO calculations. The extracted sound velocity of the lowest lying phonon modes are remarkably isotropic, suggesting that the lattice dynamics are quite three-dimensional unlike the two-dimensional nature of magnetism in \RuCl{}.
%Our extracted phonon velocities show that the in-plane phonons polarized out-of-plane and the out-of-plane phonons polarized in-plane have a much lower velocity than other branches. According to theory, the higher velocities of these modes infers higher edge-bulk coupling and thus these are the most important modes to observe the half-integer quantized thermal Hall effect \cite{yeQuantizationThermalHall2018, vinkler-avivApproximatelyQuantizedThermal2018}. The importance of these modes related to interlayer dynamics also explains the sensitivity of half-quantized thermal Hall conductivity to stacking faults.
Additionally, we performed a temperature study which found a first-order structural transition to a non-$C2/m$ space group at ${\approx}$145~K. Leveraging the high energy resolution and small beam size of IXS, our results unambiguously show the same transition in both the elastic and inelastic channels, proving the existence of this bulk transition. Besides a shift in the $\Gamma$ point across the transition, the acoustic phonon dispersion is insensitive to structure, stacking faults, and temperature, which further supports the idea of a stacking layer change rather than a local change.

Together our static and dynamic lattice studies are important benchmarks for constraining edge-bulk theories explaining the approximately half-quantized thermal Hall effect in \RuCl{}. Furthermore, our results are useful for calculating other quantities (i.e. phonon mean free path) and for subtracting phononic contributions in other techniques such as Raman spectroscopy and heat capacity, which are often subtle and difficult \cite{widmannThermodynamicEvidenceFractionalized2019}.

%%%%%%%%%%%%%%%%%%%%%%%%%%%%%%%%%%%%%%%%%%%%%%%%%%%%%%%%%%
%%%%%%%%%%%%%%%%%%%%%%%%%%%%%%%%%%%%%%%%%%%%%%%%%%%%%%%%%%
%%%%%%%%%%%%%%%%%%%%%%%%%%%%%%%%%%%%%%%%%%%%%%%%%%%%%%%%%%
%\BWL{MAT\&METH} 
%%%%%%%%%%%%%%%%%%%%%%%%%%%%%%%%%%%%%%%%%%%%%%%%%%%%%%%%%%
%%%%%%%%%%%%%%%%%%%%%%%%%%%%%%%%%%%%%%%%%%%%%%%%%%%%%%%%%%
%%%%%%%%%%%%%%%%%%%%%%%%%%%%%%%%%%%%%%%%%%%%%%%%%%%%%%%%%%

Single crystals of \RuCl{} were synthesized by vacuum sublimation \cite{plumbRuClSpinorbitAssisted2014} and their properties were characterized \cite{searsMagneticOrderRuCl2015, kimSampleQualityRuCl32021}. 200~$\mu$m-thick crystals were used to ensure high quality and reduce possible handling damage \cite{kimSampleQualityRuCl32021}. Heat capacity measurements showed a single peak at 7.2~K and x-ray diffraction showed no diffuse scattering, indicating high-quality single crystals with low stacking faults (see Supplemental Material). We use the hexagonal notation corresponding to the the trigonal $P3_112$ space group throughout this paper \cite{stroganov1957}. The lattice constants at 5.7~K were $a=b=5.96$~\AA{} and $c=17.3$~\AA{}. The corresponding Brillouin zone and high-symmetry points are shown in the inset of Fig.~\ref{fig:dispersion}.

IXS measurements were performed at the 30-ID beamline \cite{saidHighenergyresolutionInelasticXray2020, toellnerSixreflectionMeVmonochromatorSynchrotron2011} of the Advanced Photon Source. The incident beam was 35 $\upmu$m $\times$ 15 $\upmu$m with E $=$ 23.742~keV and 1.3~meV FWHM total energy resolution.

Phonon dispersions were calculated in the \textsc{phonopy} ~\cite{togoFirstPrinciplesPhonon2015} package via the frozen-phonon method. Atomic forces were extracted from DFT+$U$+SO calculations with 0.01~\AA{} atomic displacements using \textsc{vasp}~\cite{kresseEfficientIterativeSchemes1996,kresseUltrasoftPseudopotentialsProjector1999} with the Perdew-Burke-Ernzerhof (PBE) flavor of the exchange-correlation potential~\cite{perdewGeneralizedGradientApproximation1996} with Grimme's D2 correction~\cite{grimmeSemiempiricalGGAtypeDensity2006}. Calculations without a D2 correction are shown in the Supplemental Material. We used a 64-atom supercell ($2\times1\times2$) for $C2/m$ \cite{caoLowtemperatureCrystalMagnetic2016} and a 96-atom supercell ($2\times2\times1$) for $R\bar3$ \cite{parkEmergenceIsotropicKitaev2016}. All calculations in the manuscript were performed with a ferromagnetic (FM) spin configuration. Calculations with a zigzag antiferromagnetic (ZZ AFM) spin configuration \cite{searsMagneticOrderRuCl2015,chaloupkaZigzagMagneticOrder2013,parkEmergenceIsotropicKitaev2016,caoLowtemperatureCrystalMagnetic2016,johnsonMonoclinicCrystalStructure2015} showed minimum change to acoustic phonons (see Supplemental Material). The DFT+$U$+SO parameters were set to $U=1.8$ eV for the on-site Coulomb repulsion and $J=0.4$ eV for the Hund's coupling~\cite{widmannThermodynamicEvidenceFractionalized2019}. The Brillouin zone was sampled by a $5\times5\times5$ Monkhorst-Pack k-mesh.

%%%%%%%%%%%%%%%%%%%%%%%%%%%%%%%%%%%%%%%%%%%%%%%%%%%%%%%%%%
%%%%%%%%%%%%%%%%%%%%%%%%%%%%%%%%%%%%%%%%%%%%%%%%%%%%%%%%%%
%%%%%%%%%%%%%%%%%%%%%%%%%%%%%%%%%%%%%%%%%%%%%%%%%%%%%%%%%%
%\BWL{RESULTS}
%%%%%%%%%%%%%%%%%%%%%%%%%%%%%%%%%%%%%%%%%%%%%%%%%%%%%%%%%%
%%%%%%%%%%%%%%%%%%%%%%%%%%%%%%%%%%%%%%%%%%%%%%%%%%%%%%%%%%
%%%%%%%%%%%%%%%%%%%%%%%%%%%%%%%%%%%%%%%%%%%%%%%%%%%%%%%%%%

The phonon dispersion of \RuCl{} is shown in Fig.~\ref{fig:dispersion}. These measurements were performed below the N\'{e}el temperature: 4.1~K for the ZA modes and 5.7~K for all other modes. The IXS spectra and theirs fits are shown in the Supplemental Material.

The phonon modes were measured along the high-symmetry directions shown in the inset of Fig.~\ref{fig:dispersion} and are labelled in reduced wavevector notation at the bottom and as a function of high-symmetry points at the top. The scattering geometry was chosen to measure specific acoustic phonon modes according to the $\mathbf{Q}\cdot\mathbf{e}(\mathbf{q})$ term of the structure factor, where \textbf{Q} is the scattering wavevector and $\mathbf{e}(\mathbf{q})$ is the phonon polarization as a function of reduced wavevector. 

Longitudinal acoustic (LA), in-plane transverse acoustic (TA), and out-of-plane transverse acoustic (ZA) phonons are shown as orange squares, inverted dark red triangles, and purple triangles respectively, except for the purple triangles along \GA{} which represent TA phonons since the momentum is out-of-plane. Optical phonons are plotted as gray circles, but they were not systematically measured in this study. The edges of the circles are colored to indicate measurement during an LA (orange) or TA (dark red) scan.

The phonons along \GA{} disperse until (0,~0,~1.5) instead of A(0,~0,~0.5), with no sign of zone-folding. We are using the three-layer $P3_112$ space group, therefore this indicates that phonons propagating along \GA{} have single-layer periodicity in real space. 

Our calculated phonon dispersions are shown as gray lines for the $R\bar3$ (solid) and $C2/m$ (dotted) space groups. The agreement with experiment is good for both structures. One can see that certain modes are better described by one structure, but overall the trend remains the same. Furthermore, as shown in the Supplemental Material, removing the Grimme's D2 correction \cite{grimmeSemiempiricalGGAtypeDensity2006} improves the in-plane TA modes for $R\bar3$ while underestimating the in-plane LA modes. It is also shown that switching from FM to ZZ AFM has a minor effect on the acoustic phonon dispersion, but it is more dramatic for some optical phonons. Another tuning parameter is Coulomb correlations, in our case we used the rotationally invariant method of \citet{liechtensteinDensityfunctionalTheoryStrong1995} with $U=1.8~\mathrm{eV}$ and $J=0.4~\mathrm{eV}$. For comparison, \citet{muRoleThirdDimension2021} found the best model for their INS measurements with $R\bar3$, ZZ AFM, simplified rotationally invariant correlation scheme of \citet{dudarevElectronenergylossSpectraStructural1998} with $U_{\mathrm{eff}} = U - J = 3~\mathrm{eV}$, and Grimme's D3 correction \cite{grimmeConsistentAccurateInitio2010}.

%%%%%%%%%%%%%%%%%%%%%%%%%%%%%%%%%%%%%%%%%%%%%%%%%%%%%%%%%%

\begin{figure}[t]
\includegraphics{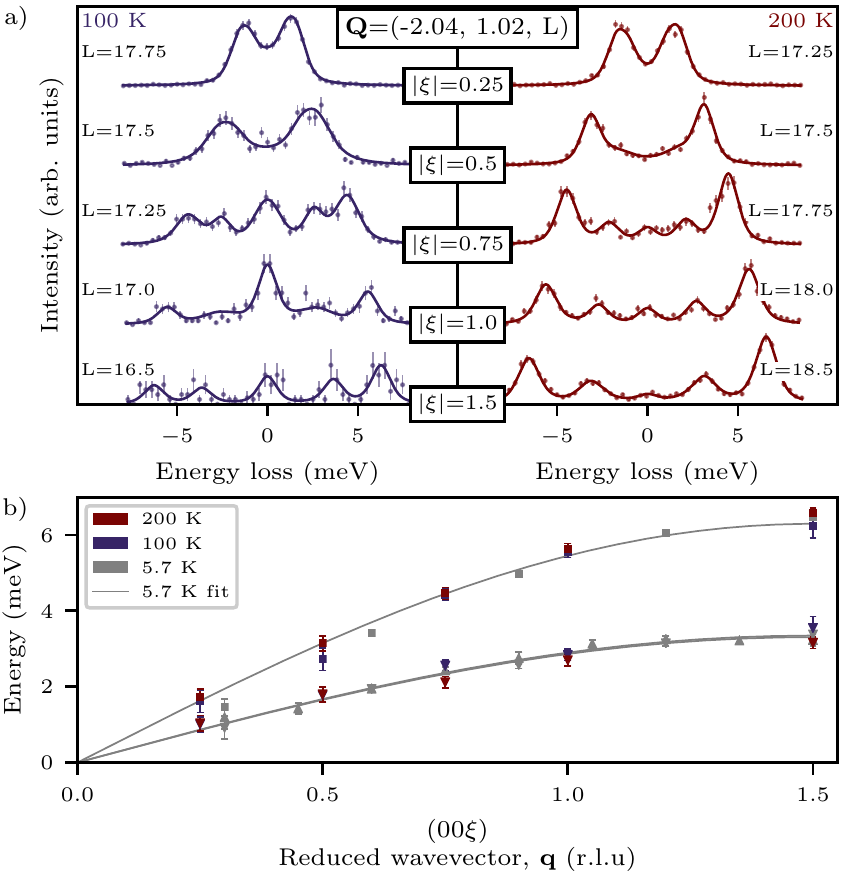}
\caption{\label{fig:transition1} (a) Inelastic x-ray scattering spectra collected below (left, 100~K) and above (right, 200~K) structural transition. Spectra were collected at \textbf{Q}=($-2.04$,~1.02,~L) which measures the TA and LA modes simultaneously along \GA{}. The $\Gamma$ point shifts from ($-2$,~1,~18) at 100~K to ($-2$,~1,~17) at 200~K. (b) TA and LA phonon dispersion extracted from spectra in panel (a) at 200~K (dark red) and 100~K (dark blue). The 5.7~K data is shown in gray along with its sine dispersion fit.}
\end{figure}

%%%%%%%%%%%%%%%%%%%%%%%%%%%%%%%%%%%%%%%%%%%%%%%%%%%%%%%%%%

Phonon velocities extracted from experimental dispersion are given in Table~\ref{tab:velocity}. The TA and LA phonon velocities are taken from linear fits near the zone center, while the ZA phonon velocities are extracted from fits of $E = \sqrt{Aq^2 + Bq^4}$ near the zone center. The out-of-plane dispersion of \RuCl{} is comparable to other van der Waals materials, such as graphite or MoS$_2$ \cite{mohrPhononDispersionGraphite2007,tornatzkyPhononDispersionMoS2019}. However, the in-plane dispersion of \RuCl{} is much softer due to its ionic rather than covalent bonds. As a result, the dispersion of \RuCl{} is much more isotropic than graphite or MoS$_2$.

%%%%%%%%%%%%%%%%%%%%%%%%%%%%%%%%%%%%%%%%%%%%%%%%%%%%%%%%%%

\begin{table}[t]
\caption{\label{tab:velocity}Phonon velocities (km/s) in \RuCl{}}
\begin{ruledtabular}
\begin{tabular}{l d d d}
 & \multicolumn{1}{c}{~~~~~\GK{}} & \multicolumn{1}{c}{~~~~~\GM{}} & \multicolumn{1}{c}{~~~~~\GA{}}\\
\hline
LA    & 4.1(2) & 4.9(4) & 2.3(3) \\
TA    & 2.6(1)  & 2.4(2)  & 1.3(1)  \\
ZA/TA & 0.9(2) & 1.0(2)  & 1.4(2) \\
\end{tabular}
\end{ruledtabular}
\end{table}

%%%%%%%%%%%%%%%%%%%%%%%%%%%%%%%%%%%%%%%%%%%%%%%%%%%%%%%%%%

The phonon dispersion along \GA{} was also explored as a function of temperature. Fig.~\ref{fig:transition1}a shows phonon spectra measured at 100~K (left) and 200~K (right). The spectra were measured at $\mathbf{Q}$ = ($-2.04$,~1.02,~L) for varying L, where the in-plane components are slightly offset from the 2D zone center to reduce the elastic background. Each spectrum contains five peaks: an elastic peak, TA and LA peaks since $\mathbf{Q}\cdot\mathbf{e}(\mathbf{q})$ has components of both, as well as the corresponding anti-Stokes phonon modes on the energy gain side of the spectrum.

\begin{figure}[t]
\includegraphics{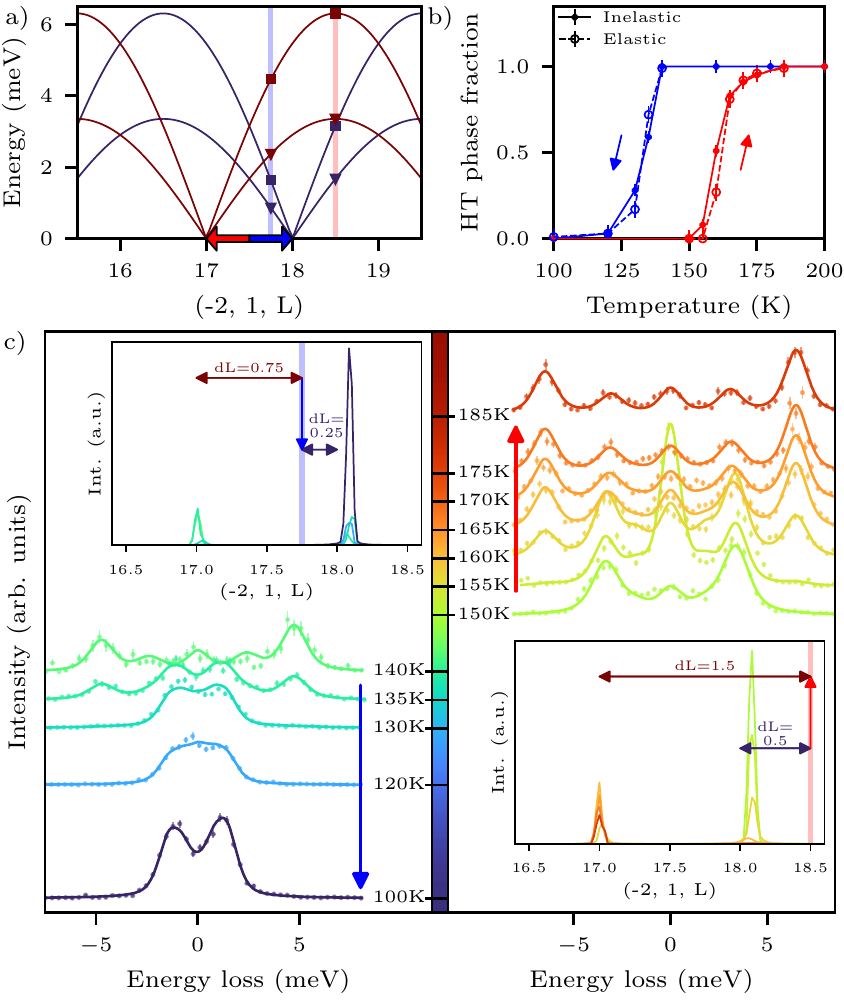}
\caption{\label{fig:transition2} Temperature study of inelastic and elastic x-ray scattering at ($-2$,~1,~L). (a) Diagram showing shift of $\Gamma$ due to structural transition, with corresponding shift of phonon dispersion. The L value used for inelastic measurements during cooling and warming are shown as light blue and red lines. Energy of LA (square) and TA (inverted triangle) phonons are shown at low (dark blue) and high (dark red) temperature. (b) High-temperature phase fraction extracted from inelastic (solid circle) and elastic (open circle) measurements as a function of cooling (blue) and warming (red). (c) Inelastic spectra during cooling (left) and warming (right) with temperature scale in middle. The insets show corresponding elastic measurements. Cooling and warming measurements were performed at ($-2.04$,~1.02,~17.75) and ($-2.04$,~1.02,~18.5) respectively, as shown by lightly colored blue (cooling) and red (warming) lines in panel (a) and insets.}
\end{figure}

The 100~K and 200~K spectra have equivalent dispersions, however they have different zone centers: ($-2$,~1,~18) at 100~K and ($-2$,~1,~17) at 200~K. This shift is shown schematically in Fig.~\ref{fig:transition2}a. The dispersions are compared directly in Fig.~\ref{fig:transition1}b, where the fitted phonon energies are plotted. The 100~K and 200~K data both show good agreement with each other, as well as the 5.7~K data and its sine dispersion fit. This zone center shift is due to a structural transition, corresponding to a change in the stacking order of the RuCl$_3$ layers. The shift from ($-2$,~1,~17) to ($-2$,~1,~18) agrees with a change from high-temperature $C2/m$ to low-temperature $R\bar3$, however further structural refinements would be needed to confirm the low-temperature space group. ($-2$,~1,~18) is also an allowed reflection of the $P3_112$ space group, however it has a vanishing structure factor.

This structural transition was further studied as a function of cooling and warming at a fixed $\mathbf{Q}$. The cooling and warming were measured at ($-2$,~1,~17.75) and ($-2$,~1,~18.5) respectively, as shown by the light blue and red vertical lines in Fig.~\ref{fig:transition2}a\&c. The dark red and blue markers on the vertical lines in Fig.~\ref{fig:transition2}a show the expected phonon energies at high and low temperature respectively, i.e. we can imagine Fig.~\ref{fig:transition2}c as the cuts of Fig.~\ref{fig:transition2}a with additional anti-Stokes peaks at negative energy loss. For example: when cooling we see the blue line shows the cut and if we are at high temperature the red dark symbols show us we expect peaks at 2.4~meV (triangle) and 4.5~meV (square) which we see on the left panel of Fig.~\ref{fig:transition2}c for the 140~K spectrum.

Focusing on the cooling, we can see the IXS spectra on the left side of Fig.~\ref{fig:transition2}c, along with diffraction data along ($-2$,~1,~L) in the inset. The elastic data shows the L=17 peak gradually disappears and is replaced by the L=18 peak. We see the same coexistence with the IXS spectra, with two distinct peaks (dL=0.75) becoming a broad peak due to the two unresolved phonons being close in energy towards $\Gamma$ (dL=0.25). Similar conclusions are drawn from the warming data on the right of Fig.~\ref{fig:transition2}c.

As shown previously in Fig.~\ref{fig:transition1}, the LA and TA phonon dispersions along \GA{} show very little temperature dependence. Therefore, we used this dispersion to constrain the fits (solid lines in Fig.~\ref{fig:transition2}c) of the spectra as a function of phase fraction. In Fig.~\ref{fig:transition2}b, the fitted high-temperature phase fraction is shown as a solid circles during cooling (blue) and warming (red). The phase fraction extracted from the relative integrated intensities of the elastic peaks is also plotted as open circles. 

The coexistence of two phases and lack of phonon softening proves that the structural phase transition is first-order. The small beam and high energy resolution of IXS helps us conclude that the space group of the low-temperature phase is not $C2/m$ and that it is likely $R\bar3$. The transition temperature (145~K) and width (30~K) agrees well with previous reports using a variety of probes \cite{kubotaSuccessiveMagneticPhase2015,searsNeutronXrayDiffraction2017,parkEmergenceIsotropicKitaev2016, widmannThermodynamicEvidenceFractionalized2019, gassFieldinducedTransitionsKitaev2020, reschkeElectronicPhononExcitations2017, reschkeSubgapOpticalResponse2018, glamazdaRelationKitaevMagnetism2017, wangMagneticExcitationsContinuum2017, heUniaxialHydrostaticPressure2018}. Previous results with larger hysteresis may be due to sample quality or kinetics (our measurements were a few hours per point, much slower than some probes).
%%%%%%%%%%%%%%%%%%%%%%%%%%%%%%%%%%%%%%%%%%%%%%%%%%%%%%%%%%

%%%%%%%%%%%%%%%%%%%%%%%%%%%%%%%%%%%%%%%%%%%%%%%%%%%%%%%%%%

%%%%%%%%%%%%%%%%%%%%%%%%%%%%%%%%%%%%%%%%%%%%%%%%%%%%%%%%%%
%%%%%%%%%%%%%%%%%%%%%%%%%%%%%%%%%%%%%%%%%%%%%%%%%%%%%%%%%%
%%%%%%%%%%%%%%%%%%%%%%%%%%%%%%%%%%%%%%%%%%%%%%%%%%%%%%%%%%
%\BWL{DISCUSSION}
%%%%%%%%%%%%%%%%%%%%%%%%%%%%%%%%%%%%%%%%%%%%%%%%%%%%%%%%%%
%%%%%%%%%%%%%%%%%%%%%%%%%%%%%%%%%%%%%%%%%%%%%%%%%%%%%%%%%%
%%%%%%%%%%%%%%%%%%%%%%%%%%%%%%%%%%%%%%%%%%%%%%%%%%%%%%%%%%

Overall, our results show a fairly regular and isotropic acoustic phonon dispersion in \RuCl{} which agrees with our frozen-phonon calculations. Furthermore, we demonstrate unequivocal evidence of a first-order structural transition upon lowering temperature which corresponds to a change of stacking order.

The dispersion in Fig.\ref{fig:dispersion} was measured below the N\'{e}el temperature, however measurements from 5~K to 300~K in four different runs showed no major deviation in phonon velocity for the ZA\GK{}, ZA\GM{}, and TA\GA{} modes (see Supplementary Information). These are the phonons with the lowest velocity, around 1 km/s (see Table \ref{tab:velocity}). \citet{yeQuantizationThermalHall2018} showed $\gamma \propto v^{-4}$, where $\gamma$ is the edge-bulk coupling and $v$ is the phonon velocity. Therefore, these low-velocity phonons are the most important for determining $\gamma$, which must be sufficiently strong for quantization in the scheme of \citet{yeQuantizationThermalHall2018}. Comparison of our measurements on different runs and samples (see Supplementary Information) also indicates very little sample dependence, i.e. stacking faults do not modify the phonon dispersion. 

Our extracted phonon velocities agree quite well with INS measurements from \citet{muRoleThirdDimension2021}, in particular for the low energy phonons of interest: we find in-plane ZA and out-of-plane TA modes in a 0.9--1.4~km/s range compared to their 1.10--1.43~km/s range. The largest discrepancy is for the LA\GA{} mode where we find 2.3~km/s versus their 3.11~km/s. This difference is reasonable, considering our calculations in Fig.~\ref{fig:dispersion} and Supplemental Material show this mode consistently overestimated in calculations, which \citet{muRoleThirdDimension2021} are relying on for extracting phonon velocities. Our results also agree fairly well with \citet{liDivergenceMajoranaPhononScattering2021}, although there is relatively large discrepancy between the LA\GK{} modes. Their velocities were 3.5 km/s, 2.5 km/s, and 1.7 km/s for the LA\GK{}, TA\GM{}, and TA\GA{} modes respectively (1~km/s $=$ 6.58~meV$\cdot$\AA, since $2\pi$ factor included in their $q$). \citet{liDivergenceMajoranaPhononScattering2021} also observe three-layer periodicity for the TA\GA{} mode, in contrast to the single-layer periodicity of our results and \citet{muRoleThirdDimension2021}. 

Our results do not show any anomalies due to coupling with excitations from 5--300~K. Previous IXS phonon measurements of the transverse phonon along \GM{} found two low-energy interlaced phonons between 3--7~meV \cite{liGiantPhononAnomalies2021}, which are not visible in our results. It is important to note that they measured these phonons along (6,-3,0)-(6,-2,0), however (6,-2,0) is not a true $\Gamma$ point. Our calculations reproduce one of the modes extremely well if we follow the same path, however we stress that this path does not pass through the M point (see Supplemental Material).

The applicability of our zero field results to the half-integer quantized thermal Hall effect should be discussed since this occurs at a field of $\approx$8~T. The minor difference of acoustic phonon dispersion calculations with FM and ZZ AFM configurations (see Supplemental Material) suggests a minor role for magnetism and that our experimental results can be extrapolated to the QSL region. Future high-field measurements would be desirable.

%%%%%%%%%%%%%%%%%%%%%%%%%%%%%%%%%%%%%%%%%%%%%%%%%%%%%%%%%%
%%%%%%%%%%%%%%%%%%%%%%%%%%%%%%%%%%%%%%%%%%%%%%%%%%%%%%%%%%
%%%%%%%%%%%%%%%%%%%%%%%%%%%%%%%%%%%%%%%%%%%%%%%%%%%%%%%%%%
%\BWL{CONCLUSION}
%%%%%%%%%%%%%%%%%%%%%%%%%%%%%%%%%%%%%%%%%%%%%%%%%%%%%%%%%%
%%%%%%%%%%%%%%%%%%%%%%%%%%%%%%%%%%%%%%%%%%%%%%%%%%%%%%%%%%
%%%%%%%%%%%%%%%%%%%%%%%%%%%%%%%%%%%%%%%%%%%%%%%%%%%%%%%%%%

In conclusion, we measured the acoustic phonon dispersion in \RuCl{} which agrees very well with our frozen-phonon calculations. We also show conclusively a first-order structural transition at ${\approx}$145~K to a non-$C2/m$ space group. These results are integral to the understanding of \RuCl{}, in particular the recently observed half-quantized thermal Hall effect. 

%%%%%%%%%%%%%%%%%%%%%%%%%%%%%%%%%%%%%%%%%%%%%%%%%%%%%%%%%%
%%%%%%%%%%%%%%%%%%%%%%%%%%%%%%%%%%%%%%%%%%%%%%%%%%%%%%%%%%
%%%%%%%%%%%%%%%%%%%%%%%%%%%%%%%%%%%%%%%%%%%%%%%%%%%%%%%%%%

\begin{acknowledgments}
Work at the University of Toronto was supported by the Natural Science and Engineering Research Council (NSERC) of Canada, Canadian Foundation for Innovation, and Ontario Innovation Trust. This research used resources of the Advanced Photon Source (APS), a DOE Office of Science User Facility operated for the DOE Office of Science by Argonne National Laboratory under Contract No. DE-AC02-06CH11357. B.W.L. acknowledges support from the University of Toronto Faculty of Arts and Sciences Postdoctoral Fellowship and NSERC [funding reference number PDF-546035-2020]. The work of D.A.P. was supported by the Russian Science Foundation, Grant No. 21-72-10136. The work in Augsburg was supported by the German Research Foundation (DFG) via the Project No. 107745057 (TRR80). 
\end{acknowledgments}

%%%%%%%%%%%%%%%%%%%%%%%%%%%%%%%%%%%%%%%%%%%%%%%%%%%%%%%%%%
%%%%%%%%%%%%%%%%%%%%%%%%%%%%%%%%%%%%%%%%%%%%%%%%%%%%%%%%%%
%%%%%%%%%%%%%%%%%%%%%%%%%%%%%%%%%%%%%%%%%%%%%%%%%%%%%%%%%%

\bibliography{main}
\end{document}